\documentclass{PoS}
\usepackage[latin1]{inputenc}
\usepackage[T1]{fontenc}

\usepackage{amsmath}
\usepackage{amssymb}

\title{Infrared Behavior of Three-Point Functions in Landau Gauge Yang-Mills Theory}

\ShortTitle{Infrared Behavior of Three-Point Functions in Landau Gauge Yang-Mills Theory}

\author{\speaker{Markus Huber}%
\\
        Karl-Franzens University Graz\\
        E-mail: \email{markus.huber@uni-graz.at}}

\author{Reinhard Alkofer\\
       Karl-Franzens University Graz\\
       E-mail: \email{reinhard.alkofer@uni-graz.at}}

\author{Kai Schwenzer\\
       Karl-Franzens University Graz\\
       E-mail: \email{kai.schwenzer@uni-graz.at}}

\abstract{The three-gluon and ghost-gluon vertices of Landau gauge Yang-Mills theory are investigated in the low momentum regime. Due to ghost dominance in the infrared we can use the known power law behavior for the propagators to determine analytically the complete momentum dependence of the dressing functions. Besides a uniform, i.~e. all momenta going to zero, divergence, we find additional singularities, if one momentum alone goes to zero, while the other two remain constant. At these asymmetric points we can extract additional infrared exponents, which corroborate previous results and expand the known fixed point solution of Landau gauge Yang-Mills theory, where the uniform infrared exponents for all vertex functions are known.
Calculations in two and three dimensions yield qualitatively similar results.

We find several dressing functions diverging like $(p^2)^{1-2\ka}$, if only the gluon momentum $p$ goes to zero. Of these many are longitudinal and do not contribute to Dyson-Schwinger equations. The divergent transversal parts are additionally suppressed by the corresponding tensor. The longitudinal dressing function of the ghost-gluon vertex behaves similarly, when the gluon momentum becomes small compared to the ghost momentum, whereas all its other dressing functions vanish at the asymmetric points. The uniform momentum dependence of the three-gluon vertex is determined as $(p^2)^{-3\ka}$, while the ghost-gluon vertex stays finite in this limit.
}

\FullConference{8th Conference Quark Confinement and the Hadron Spectrum\\
		 September 1-6, 2008\\
		 Mainz. Germany}

\newcommand{\mhalf}[1]{\frac{#1}{2}}
\newcommand{\ka}{\kappa}
\newcommand{\fref}[1]{fig.~\ref{#1}}
\newcommand{\nq}{\nu_1}
\newcommand{\nw}{\nu_2}
\newcommand{\nd}{\nu_3}

\begin{document}

\section{Infrared Behavior}

Many non-perturbative aspects of Yang-Mills theory are encoded in the infrared (IR) behavior of its Green functions. For very small momenta below the intrinsic scale $\Lambda_{QCD}$ they can be described by power laws. This scaling solution \cite{von Smekal:1997is} is in accordance with the scenarios for confinement of Gribov-Zwanziger and Kugo-Ojima. The general expression for the so-called IR exponent of an arbitrary Green function with $m$ gluon and $2n$ ghost legs under the assumption that all momenta go to zero uniformly is \cite{Alkofer:2004it}
\begin{equation}
\label{eq:ir-exp-dominant}
\delta_{2n,m}=(n-m)\ka+(1-n)\left(\mhalf{d}-2\right).
\end{equation}
This formula is valid in $d=2,3,4$ dimensions for the corresponding values of $\kappa$. As it is clear that we can have more than one independent momentum for vertex functions, the question arises what happens when only one of these goes to zero? Interestingly it turns out that additional divergences can occur that do not change the uniform solution \cite{Alkofer:2008jy}.

For the calculation of the three-point vertices we used the known power laws for the propagators and the following truncations for the Dyson-Schwinger equations (DSEs), which is motivated by ghost dominance in the IR \cite{von Smekal:1997is,Zwanziger:1993dh}:


 \includegraphics[width=0.45\linewidth]{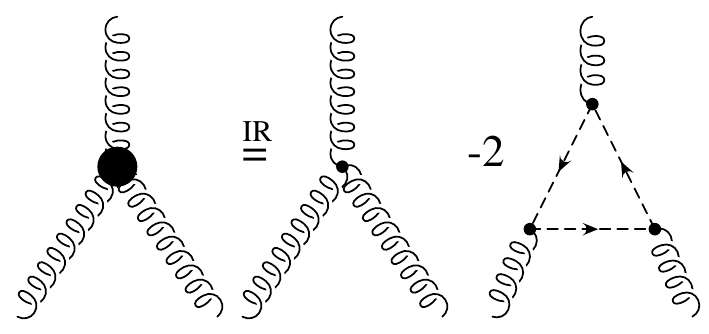}, $\quad\quad$
 \includegraphics[width=0.45\linewidth]{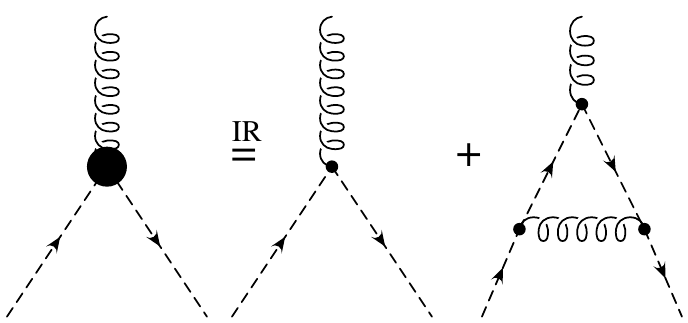}.

For the dressed ghost-gluon vertex we used the bare one, justified by a simple argument of Taylor \cite{Taylor:1971ff}, which is supported by lattice simulations \cite{Cucchieri:2008} and the DSE solution \cite{Schleifenbaum:2004id}. The truncation considers only the first order of the skeleton expansion. For more details see \cite{Alkofer:2008dt}.

\section{Three-point integral}

The ghost-triangle diagram of the three-gluon vertex is decomposed in the IR into ten tensors $\tau^i_{\mu \nu \rho}(p_1,p_2,p_3)$ and ten corresponding dressing functions $E_i(p_1,p_2,p_3)$.
The applied tensor decomposition \cite{Davydychev:1991va} reveals that only ten instead of the expected fourteen scalar functions $E_i$ and tensors $\tau^i_{\mu \nu \rho}$ are necessary. The former consist of massless three-point integrals,
\begin{equation}
\int \frac{d^{d}q}{(2\pi)^{d}} \frac{1} 
		{((q+p_1)^2)^{\boldsymbol{\nq}}((q-p_2)^2)^{\boldsymbol{\nw}}(q^2)^{\boldsymbol{\nd}}},
\end{equation}
where $\nu_1$, $\nu_2$ and $\nu_3$ are non-integer numbers. We employed the Negative Dimensions Integration Method (NDIM) which yields a full analytic solution in terms of Appell's series $F_4$. Using several different analytic continuations we can calculate the momentum dependence of the diagram for arbitrary $d$. As the variables of the Appell's series are the momentum ratios $p_1^2/p_3^2$ and $p_2^2/p_3^2$ we plot the scalars $E_i$ from the tensor decomposition as functions of $p_1^2$ and $p_2^2$ with $p_3^2$ fixed. The plotted region in \fref{fig:kinSing-E10} represents the accessible ratios for Euclidean momenta and is restricted by momentum conservation. The ghost-gluon vertex can be decomposed similarly into two parts.

\section{Three-point vertex functions}

\begin{figure}[t]
 \includegraphics[width=0.36\linewidth]{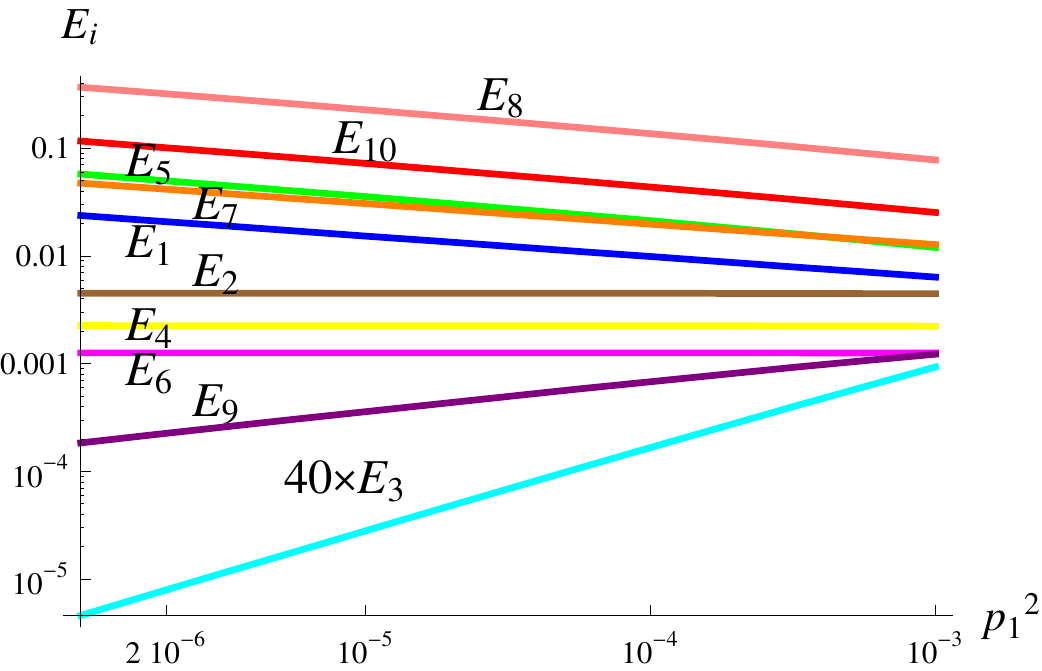}
 \includegraphics[width=0.28\linewidth]{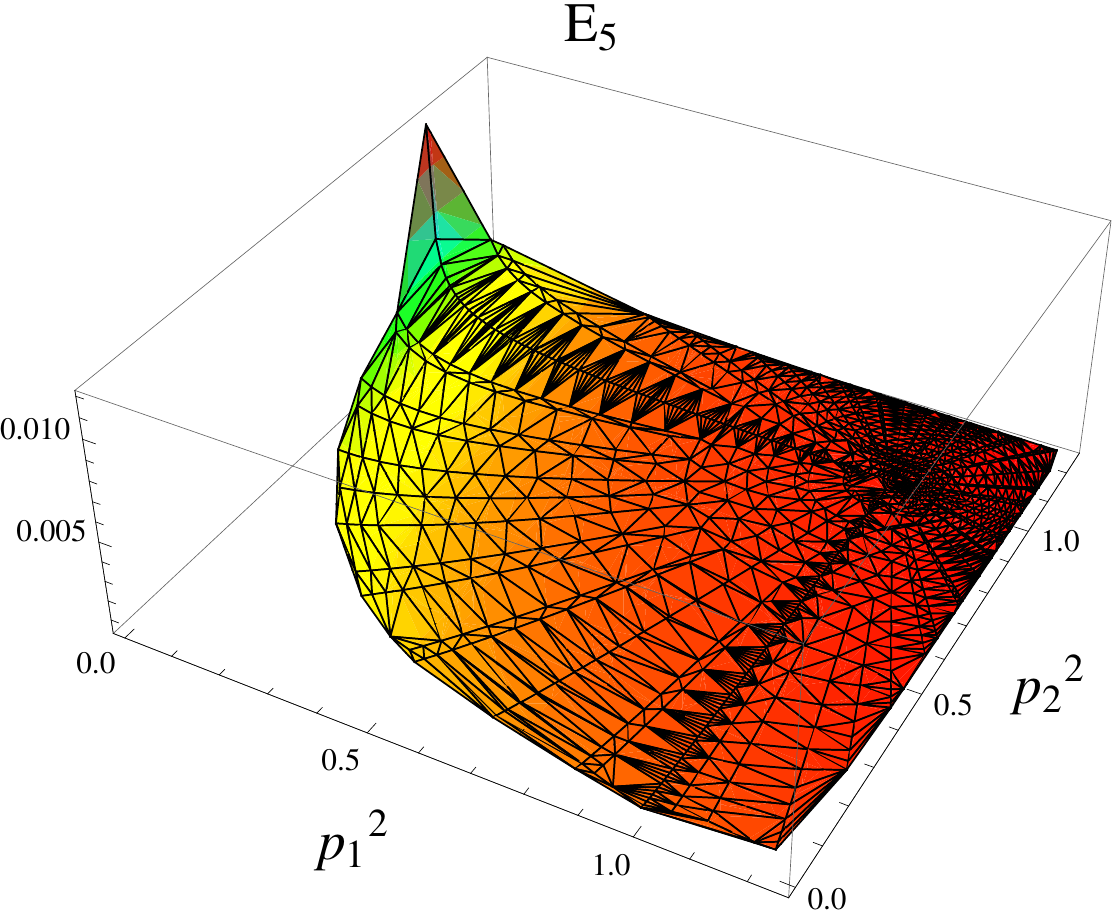}
\includegraphics[width=0.28\linewidth]{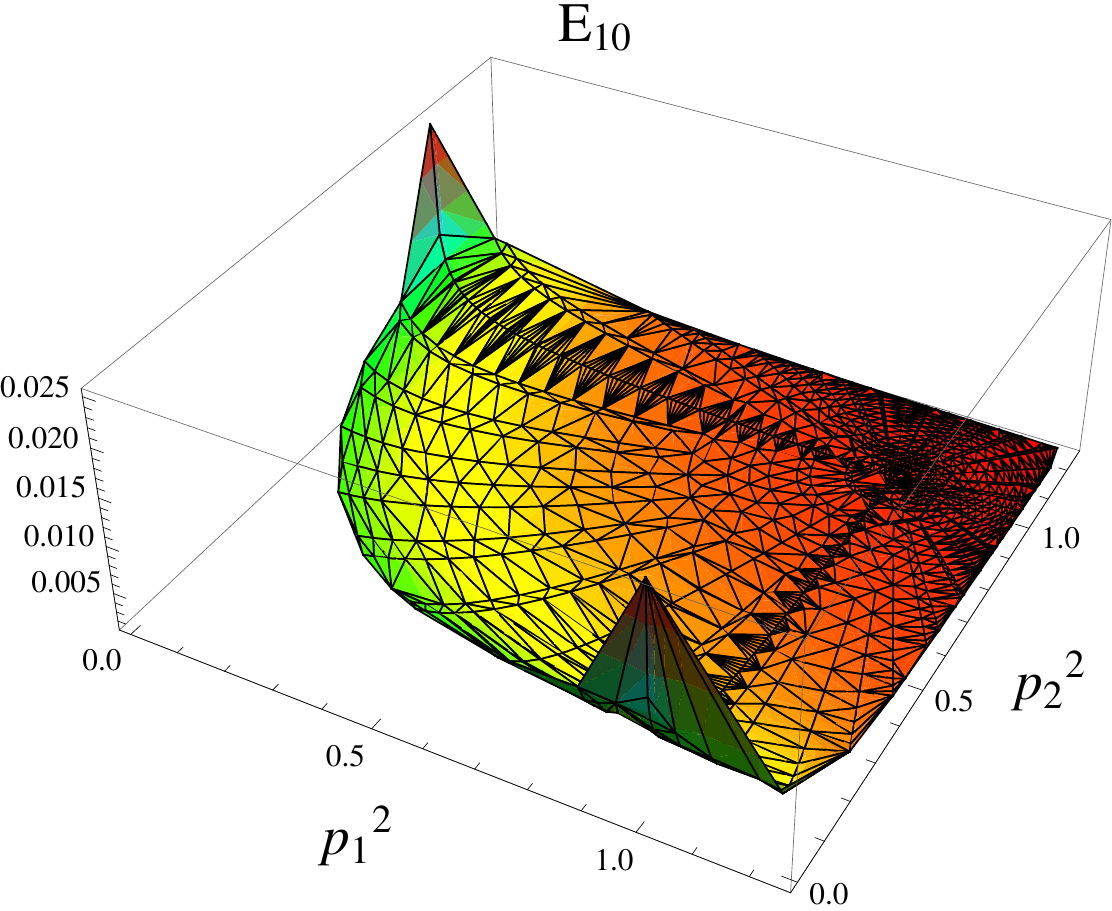}
\caption{Left: A double-logarithmic plot of the scalar functions $E_i$ of the three-gluon vertex when $p_1^2$ becomes small and $p_3^2=p_2^2$ is held fixed. Three cases can be distinguished: Five scalar functions diverge like $(p_1^2)^{1-2\ka}$, three stay constant, two vanish like $(p_1^2)^{3/2-2\ka}$ and $(p_1^2)^{2-2\ka}$ respectively.
Middle and right: The scalar functions $E_5$ and $E_{10}$ with $p_3^2=1$. The kinematic singularities at the asymmetric points, where $p_1^2$ or $p_2^2$ go to zero, clearly dominate the structure.
}
\label{fig:kinSing-E10}
\end{figure}

As an example for the kinematic dependence of the dressing functions we show $E_5$ and $E_{10}$ in \fref{fig:kinSing-E10}.
The behavior when one of the three momenta gets small compared to the others, i.e. the region around the asymmetric points $(0,1)$, $(1,0)$ and $(\infty,\infty)$, is of special interest. This case corresponds to the emission/absorption of a soft gluon. We extracted the exponents in $d$ dimensions and indeed found kinematic singularities, which are in agreement with the power counting analysis in ref. \cite{Alkofer:2008jy}. In \fref{fig:kinSing-E10} the results in four dimensions are given.

%

To verify the self-consistency of the assumption of a bare ghost-gluon vertex we calculated its momentum dependence for uniform scaling and the case of only one small momentum. The results for the former clearly support the bare version, whereas those for the latter show that the structure of the ghost-gluon vertex is richer than expected and features a divergence ($1-2\ka$) for the longitudinal scalar function when the gluon momentum vanishes. However, this is no contradiction to Taylor's argument because this scalar function is multiplied with the soft momentum and the divergence is suppressed.

\end{document}